\documentclass[preprint,superscriptaddress]{revtex4}

\usepackage{latexsym} 
\usepackage{amssymb}  
\usepackage{amsbsy}   
\usepackage[dvips]{color}
\usepackage{bm}   
\usepackage{mathdots}
\def\RR{\mathbb{R}}

\def\CC{\mathbb{C}}

\newcommand{\<}{\langle}
\renewcommand{\>}{\rangle} 
\newcommand{\txt}{\textstyle}
\newcommand{\dsp}{\displaystyle}
\newcommand\eqn[1]{(\ref{#1})}      
\newcommand\Eqn[1]{Eq.~(\ref{#1})}  
\newcommand{\e}{{\rm e}}   
\newcommand{\ri}{{\rm i}}
\newcommand{\rU}{{\rm U}}
\newcommand{\ru}{{\rm u}}

\newcommand{\rGL}{{\rm GL}}
\newcommand{\one}{{\openone}}

\newcommand{\half} {{\txt \frac{1}{2}}}

\newcommand{\cC}{\ensuremath{\mathcal{C}}}
\newcommand{\cP}{\ensuremath{\mathcal{P}}}
\newcommand{\cT}{\ensuremath{\mathcal{T}}}

\newcommand{\cN}{\ensuremath{\mathcal{N}}}
\newcommand{\cQ}{\ensuremath{\mathcal{Q}}}
\newcommand{\tP}{\ensuremath{\tilde\mathcal{P}}}
\newcommand{\tH}{\ensuremath{\tilde H}}
\newcommand{\tW}{\ensuremath{\tilde W}}
\newcommand{\bP}{\ensuremath{\bar\mathcal{P}}}
\newcommand{\bH}{\ensuremath{\bar H}}
\newcommand{\bW}{\ensuremath{\bar W}}

\newcommand{\Tr}{\mbox{Tr}}

\newcommand{\nn}{\nonumber \\}
\newcommand{\be}[1]{\begin{equation}\label{#1}}
\newcommand{\ee}{\end{equation}}
\newcommand{\ba}[1]{\begin{eqnarray}\label{#1}}
\newcommand{\ea}{\end{eqnarray}}
\newcommand{\rf}[1]{(\ref{#1})}
\newcommand{\dd}{\dagger}

\begin{document}

\title{\bf General \cP\cT-Symmetric Matrices}
\author{Jia-wen Deng}
\affiliation{Department of Physics, National University of Singapore, 117542, Singapore}
\author{Uwe G\"unther}
\affiliation{Helmholtz-Zentrum Dresden-Rossendorf, Postfach 510119, D-01314 Dresden, Germany}
\author{Qing-hai Wang}
\affiliation{Department of Physics, National University of Singapore, 117542,
Singapore}

\date{\today}

\begin{abstract}
  Three ways of constructing a non-Hermitian matrix with possible all real eigenvalues are discussed. They are \cP\cT\ symmetry, pseudo-Hermiticity, and generalized \cP\cT\ symmetry. Parameter counting is provided for each class. All three classes of matrices have more real parameters than a Hermitian matrix with the same dimension. The generalized \cP\cT-symmetric matrices are most general among the three. All self-adjoint matrices process a generalized \cP\cT\ symmetry. For a given matrix, it can be both \cP\cT-symmetric and \tP-pseudo-Hermitian with respect to some \tP\ operators. The relation between corresponding \cP\ and \tP\ operators is established. The Jordan block structures of each class are discussed. Explicit examples in $2\times 2$ are shown.
\end{abstract}

\begin{titlepage}
\maketitle
\renewcommand{\thepage}{}          
\end{titlepage}

\section{Introduction}

Since Bender and Boettcher published their seminal paper in 1998 \cite{BB}, the field of so-called \cP\cT-symmetric quantum mechanics grows rapidly. Two review articles have been published \cite{BenderReview,AliReview}. Applications in optics are widely studied in both theoretical predictions \cite{theory1,theory2,theory3,theory4} and experimental realizations \cite{exp1,NaturePhys1,NaturePhys2,Science,Nature}.

The essential idea of this new formulation of quantum mechanics is to allow the Hamiltonian to be apparently non-Hermitian, yet the entire spectrum is still real. In finite dimensions, this means that we may represent the Hamiltonian of a {\it non-dissipative}\ quantum system by a non-Hermitian matrix. Namely, a matrix $H$ with all real eigenvalues but $H\neq H^\dag$. In this paper, we try to classify all non-Hermitian matrices with possible all real eigenvalues. We hope our results will provide a reference on \cP\cT-symmetric matrices to the community. This work can be considered as a extension with more general parity operators studied in Ref.~\cite{BMW} and in higher dimensions discussed in Ref.~\cite{WCZ}.

For a diagonalizable matrix $H$ to have all real eigenvalues, a necessary and sufficient condition is that there exist a Hermitian matrix $W=W^\dag$ such that
\be{eqn:self-adjoint}
W H=H^\dag W
\ee
and all the eigenvalues of $W$ are positive definite \cite{Jordan}. If $H$ represents the Hamiltonian of a quantum system, then $W$ is the metric operator which defines the inner-product of the corresponding Hilbert space \cite{BalBook},
\be{eqn:inner}
(\cdot,\cdot) \equiv \<\cdot | W |\cdot\>,
\ee
where $\<\cdot|=|\cdot\>^\dag$. \Eqn{eqn:self-adjoint} immediately leads to
\be{eqn:Hermitian}
(\cdot,H\cdot) = (H\cdot,\cdot).
\ee
That is, $H$ is self-adjoint. In the conventional quantum mechanics, the Hamiltonian can always be represented by a Hermitian matrix. In this case, on can always choose $W$ to be the unity matrix. And the inner-product reduces to the familiar Dirac's one between bras and kets.

In the first paper on \cP\cT-symmetric quantum mechanics, Bender and Boettcher identified the combined symmetry of \cP (parity) and \cT (time reversal) as the generalization of Hermiticity \cite{BB}. They coined the term ``\cP\cT-symmetric Hamiltonian'' for a Hamiltonian commuting with \cP\cT,
\be{eqn:PT}
[\cP\cT,H]=0.
\ee
Later, Bender {\it et al}\ realized that the metric operator in \cP\cT-symmetric quantum mechanics is non-trivial \cite{BBJ}. To show the unitary time evolution, they introduced a so-called ``\cC\ {\it operator}'' and defined a \cC\cP\cT\ inner-product. Translating to the current terminology, Bender {\it et al}\ defined the metric operator as $W=\cP\cC$.

However, the \cP\cT\ symmetry does not guarantee the Hamiltonian to have an entirely real spectrum. It only ensures a real secular equation of $H$ because
\be{eqn:PT1}
(\cP\cT)^2=\one.
\ee
This means that all the eigenvalues of $H$ either are real or form complex conjugate pairs. If all the eigenvalues are real, we call the \cP\cT\ symmetry is {\it unbroken}. Otherwise, the \cP\cT\ symmetry is spontaneously broken. Not all \cP\cT-symmetric Hamiltonians are diagonalizable either. The Jordan block structure may be found if there is degeneracy.

In Sec.~\ref{sec:PT}, we study the \cP\cT-symmetric matrices by carefully defining the parity operator. Comparing to an early study in Ref.~\cite{BMW}, our formulation here is more general and it allows us to study the non-symmetric parity and Hamiltonian operators. We will show that an $(m+n)\times(m+n)$ \cP\cT-symmetric matrix has $(m+n)^2+2mn$ real parameters. For comparison, an $(m+n)\times(m+n)$ Hermitian matrix has $(m+n)^2$ real parameters. We will also show that all possible Jordan block structure may be found in the \cP\cT-symmetric matrices.

If $H$ has complex eigenvalues or $H$ is not diagonalizable, the metric $W$ in \eqn{eqn:self-adjoint} stops being well defined: Either $W$ becomes non-Hermitian, or the eigenvalues of $W$ are no longer positive-definite. If one maintains the Hermiticity of $W$ but allows it to have negative eigenvalues, the linear space equipped with the indefinite metric is called the {\it Pontrjagin space}\ (a finite-dimensional version of the {\it Krein space}). In the literature, the Hamiltonian $H$ with an indefinite-metric is called pseudo-Hermitian \cite{AliReview}.

In Sec.~\ref{sec:P-pseudo}, we discuss the \tP-pseudo-Hermitian matrices where \tP\ plays the role of an indefinite metric. This section can be considered as an extension of the early study in Ref.~\cite{WCZ} to higher dimensions. We show that an $(m+n)\times(m+n)$ \tP-pseudo-Hermitian matrix has $(m+n)^2+2mn$ real parameters, same as in a \cP\cT-symmetric matrix in the same dimension. A Hermitian matrix can always be considered as a special case of \tP-pseudo-Hermitian matrices. And all possible Jordan block structures may be found in the \tP-pseudo-Hermitian matrices.

It is commonly believed that the \cP\cT\ symmetry is neither necessary nor sufficient condition for a Hamiltonian to have an entire real spectrum. This is not true if we generalize the \cP\cT\ symmetry to be a combined involutory operation \bP\cT. In Sec.~\ref{sec:PT-gen}, we show that these generalized \bP\cT-symmetric matrices contain all self-adjoint matrices. The generalized \bP\cT\ symmetry is the necessary condition for a matrix to have all real eigenvalues. In the case of $N\times N$, a generalized \bP\cT-symmetric matrix has $2N^2-N$ real parameters, same as in an $N\times N$ self-adjoint matrix. All the \cP\cT-symmetric matrices studied in Sec.~\ref{sec:PT} are special cases of generalized \bP\cT-symmetric matrices. If one wants to model a finite-dimensional non-dissipative quantum system in the most general way, he/she should start with a generalized \bP\cT-symmetric matrix Hamiltonian.

The explicit $2\times 2$ matrices are presented for each class. In this dimension, it is interesting to know that all three classes coincide: A \cP\cT-symmetric matrix is also \tP-pseudo-Hermitian, and it exhibits the generalize \bP\cT\ symmetry as well. Only in higher dimensions, three classes may contain different matrices. And both \cP\cT-symmetric and \tP-pseudo-Hermitian matrices are special cases of the generalized \bP\cT-symmetric ones.

In the concluding Sec.~\ref{conclusion}, we summarize the parameter counting for different types of matrices in Table \ref{tab1}.

\section{\cP\cT\ Symmetry}
\label{sec:PT}

In this section, we study \cP\cT-symmetric matrices. Let us start with the definitions of the parity operator \cP\ and the time reversal operator \cT. Since the time reversal operator is an involution and anti-linear, we define it simply as the complex conjugation,
\begin{equation}
\cT \equiv * \quad \Rightarrow \quad \cT^2=\one.
\end{equation}
That is, for a matrix (operator), $\cT A\cT = A^*$, for a vector (state), $\cT |\psi\> = |\psi\>^*$, where $*$ represents complex conjugation.

The parity is a linear operator. In finite dimensions, we define it as a matrix, \cP. As usual, we demands that the time reversal and the parity commute from each other,
\begin{equation}
[\cP,\cT]=0.
\end{equation}
This commutation relation immediately leads to that \cP\ must be real. Together with the involutory property, \cP\ satisfies
\begin{equation}
\cP = \cP^* \quad {\rm and} \quad \cP^2 = \one.
\end{equation}
Our definition of the parity operator is more general than the one studied in Ref.~\cite{BMW}, in which the parity operator is always symmetric because only (complex) symmetric matrix Hamiltonians were considered. Here, we keep \cP\ to be real but allow it take non-symmetric form. This generalization gives us more parameters in both the parity operator and the Hamiltonian.

We say that a matrix Hamiltonian $H$ is \cP\cT-symmetric if it commutes with the combination of \cP\cT. With our definitions of \cP\ and \cT\ here, the \cP\cT-symmetric condition is
\begin{equation}
[\cP\cT,H]=0 \quad \Leftrightarrow\quad \cP H = H^* \cP .
\end{equation}
Because both \cP\ and \cT\ are involutions and they commute, the combination of $\cP\cT$\ is also an involution, $(\cP\cT)^2=\one.$ One consequence of this result is that the eigenvalues of $H$ can either be real or form complex conjugate pairs. The \cP\cT\ symmetry is not a sufficient condition for a matrix to process only real eigenvalues. If a \cP\cT-symmetric matrix Hamiltonian $H$ has an entirely real spectrum, we call that the \cP\cT\ symmetry is {\it unbroken}. In this case, all the eigenstates of $H$ are also eigenstates of \cP\cT.
\be{eqn:PTeigen}
H|E_n\> = E_n |E_n\>,\quad \Rightarrow \quad \cP\cT|E_n\> \equiv \cP|E_n\>^* = \lambda_n |E_n\>
\ee
with $E_n$ to be real $E_n=E_n^*$ and $\lambda_n$ to have modulus one, $\lambda_n = \e^{\ri\alpha_n}$. One can always choose the phase of the eigenvectors such that $\lambda_n=1$. When complex eigenvalues appear in $H$, even though $H$ is still \cP\cT-symmetric, the eigenstates of $H$ are no longer eigenstates of $\cP\cT$. In this sense, we call that the \cP\cT\ symmetry is spontaneously broken.

\subsection{Similarity transformations}

Because \cP\ is an involution, its eigenvalues can take only the values of $\pm 1.$ We may use the number of positive and negative eigenvalues to classify a finite-dimensional parity operator. Let us denote an $m+n$-dimensional parity operator with $m$ positive eigenvalues and $n$ negative eigenvalues as $\cP(m,n).$ All $\cP(m,n)$ operators with the same $m$ and $n$ can be linked by a similarity transformation. Since \cP\ is a real matrix, we consider only the real transformations. All these transformations form a general linear group,
\begin{equation}
\cP(m,n) = R \cP_0(m,n) R^{-1}, \quad {\rm with} \quad R\in  \rGL(m+n,\mathbb{R}).
\label{eqn:Pmn}
\end{equation}
This transformation maintains both reality and involutory properties of \cP. There are $(m+n)^2$ real parameters in a transformation matrix $R$, but not all of them enter \cP\ as we will see later.

The same transformation preserves \cP\cT\ symmetry. That is, if $H_0$ is $\cP_0\cT$-symmetric, then
\begin{equation}
H=R H_0 R^{-1}
\label{eqn:PT-H}
\end{equation}
is $\cP\cT$-symmetric:
\begin{equation}
\cP_0 H_0 = H_0^* \cP_0 \qquad \Leftrightarrow \qquad \cP H = H^* \cP.
\end{equation}

Just as shown in Ref.~\cite{BMW}, we may use this property to construct \cP\cT-symmetric matrices. Without loss of generality, we choose the parity operator $\cP_0$ to be diagonal,
\begin{equation}
\cP_0(m,n) = {\rm Diag}\{1,\cdots,1,-1,\cdots,-1\} = \left(
\begin{array}{cc}
\one_{m\times m} & 0\\
0 & -\one_{n\times n}
\end{array}
\right).
\end{equation}
It is easy to find that for an $(m+n)\times(m+n)$ matrix $H_0$ to be $\cP_0\cT$-symmetric, it must have the block form
\begin{equation}
H_0 = \left(
\begin{array}{cc}
A_{m\times m} & \ri B_{m\times n}\\
\ri C_{n\times m} & D_{n\times n}
\end{array}
\right),
\end{equation}
where $A$, $B$, $C$, and $D$ are all real matrices. There are $m^2+n^2+2mn=(m+n)^2$ real parameters in $H_0$.

To count the number of real parameters in a generic $(m,n)$ parity operator, let us take a closer look at the similarity transformation in \eqn{eqn:Pmn}. Many members of ${\rm GL}(m+n,\mathbb{R})$ leave $\cP_0(m, n)$ invariant. By group theory, all of such transformations form a group, it is called {\it the stabilizer subgroup} or {\it the little group}. Suppose that $R'$ is a member of this subgroup,
\be{Q}
R'\cP_0(m,n)R'^{-1}=\cP_0(m,n), \quad \Leftrightarrow \quad [R',\cP_0(m,n)]=0.
\ee
It is easy to see that $R'$ must be block diagonal,
\be{Qblock}
R'=
\left(
\begin{array}{cc}
X_{m\times m} & 0\\
0 & Y_{n\times n}
\end{array}
\right)
\ee
with $X$ and $Y$ to be real invertible matrices. Hence, the little group of $\cP_0(m,n)$ is the direct sum of  $\rGL(m,\mathbb{R})$ and $\rGL(n,\mathbb{R})$. All the nontrivial similarity transformations which defines \cP\ belong to the coset of
\be{coset}
\frac{\rGL(m+n,\mathbb{R})}{\rGL(m,\mathbb{R}) \times \rGL(n,\mathbb{R})}.
\ee
Thus, among $(m+n)^2$ real parameters in $\rGL(m+n,\mathbb{R})$ group, $m^2$ in $\rGL(m,\mathbb{R})$ subgroup and $n^2$ in $\rGL(n,\mathbb{R})$ subgroup do not enter \cP. There are $(m+n)^2-m^2-n^2=2mn$ real parameters in \cP. Combining with the parameters in $H_0$, we conclude that a \cP\cT-symmetric Hamiltonian has $(m+n)^2+2mn$ real parameters.

The metric operator can be found by solving the self-adjoint condition $W_0 H_0 = H_0^\dag W_0.$ When the parity is transformed to $\cP=R \cP_0 R^{-1}$ and the Hamiltonian changes to $H=R H_0 R^{-1}$, the metric operator transforms accordingly,
\begin{equation}
W=\left(R^{-1}\right)^\dag W_0 R^{-1}.
\end{equation}
Obviously, the transformation preserves the Hermiticity of the metric operator. That is, $W=W^\dag$ if $W_0 = W_0^\dag$. Although the eigenvalues of $W$ may change under such a transformation, but they remain to be positive.

\subsection{Jordan blocks}

Not all \cP\cT-symmetric matrices are diagonalizable. They  may form Jordan blocks. Let us denote $J_n(\lambda)$ as the $n$-dimensional Jordan block with eigenvalue $\lambda$,
\ba{eqn:Jn}
J_n(\lambda)=\lambda \one_{n\times n} + N_n=\left(
      \begin{array}{ccccc}
        \lambda & 1 & \cdots & 0 & 0 \\
        0 & \lambda & \cdots & 0 & 0 \\
        \vdots & \vdots & \ddots & \vdots & \vdots \\
        0 & 0 & \cdots & \lambda & 1\\
        0 & 0 & \cdots & 0 & \lambda \\
      \end{array}
    \right)_{n\times n}
\ea
with $N_n$ the nilpotent matrix with units on the first upper diagonal and $(N_n)^{n-1}\neq 0$, $(N_n)^n=0$. For an $N$-dimensional matrix $H$, it is always similar to a direct sum of a series of Jordan blocks by a similarity transformation $\Lambda$,
\be{eqn:Jordan}
\Lambda H \Lambda^{-1} = J_{n_1}(\lambda_1)\oplus J_{n_2}(\lambda_2)\oplus \cdots, \quad {\rm with} \quad N=n_1+n_2+\cdots.
\ee
Only for the special case of $n_1=n_2=\cdots=1$, $H$ is diagonalizable.

Since similarity transformations do not effect the diagonalizability, we only need to study the possible Jordan block structure in $H_0$. By choosing parameters in $H_0$, all possible Jordan blocks emerge in \cP\cT-symmetric matrices. For example, if we choose
\ba{}
A &=& J_m(\lambda), \nn
B &= &\left(
      \begin{array}{cccc}
        0 & 0 & \cdots  & 0 \\
       \vdots & \vdots & \cdots & \vdots \\
        0 & 0 & \cdots & 0\\
        1 & 0 & \cdots & 0 \\
      \end{array}
    \right)_{m\times n}, \nn
C &=& 0, \nn
D &=& J_n(\lambda),
\ea
then $H_0$ is similar to $J_{m+n}(\lambda)$:
\be{}
\Lambda H_0 \Lambda^{-1} = J_{m+n}(\lambda),
\ee
with
\be{}
\Lambda = \left(
\begin{array}{cc}
\one_{m\times m} & 0\\
0 & \ri \one_{n\times n}
\end{array}
\right).
\ee
Similarly, all other possible Jordan block structures in \eqn{eqn:Jordan} can be constructed by properly choosing the real matrices $A$, $B$, $C$, and $D$ in $H_0$. Of course, the construction here is only a sufficient condition to form Jordan blocks, it is not necessary.

\subsection{$2\times 2$}

Now let us show the explicit matrices in the case of $2\times 2$. We start with a simple diagonal parity operator,
\begin{equation}
\cP_0 = \sigma_3
=\left(
\begin{array}{cc}
1 & 0\\
0 & -1
\end{array}
\right),
\end{equation}
where $\sigma_3$ is the third Pauli matrix. A $\cP_0\cT$-symmetric matrix Hamiltonian can be parameterized as
\ba{eqn:P0T-H}
H_0 &=& e \sigma_0 + (\ri \rho, \gamma \sin\delta, \gamma\cos\delta)\cdot \bm{\sigma} \nn
&=&
\left(
\begin{array}{cc}
e+\gamma \cos\delta & - \ri (\gamma \sin\delta - \rho)\\
 \ri (\gamma \sin\delta + \rho) & e- \gamma \cos\delta
\end{array}
\right)
\ea
with four real parameters: $e$, $\gamma$, $\rho$, and $\delta$, where $\bm{\sigma} = (\sigma_1,\sigma_2,\sigma_3)$ are Pauli matrices and $\sigma_0=\one_{2\times 2}$. The eigenvalues of $H_0$ are
\begin{equation}
E_\pm = e \pm \sqrt{\gamma^2 - \rho^2}.
\end{equation}
They are real when $\gamma^2 \ge \rho^2$. It is the condition for \cP\cT\ symmetry not broken. The corresponding eigenvectors are
\begin{equation}
|E_\pm\> = n_\pm \left(
\begin{array}{c}
\gamma \e^{\ri\delta} -\ri \rho \pm \sqrt{\gamma^2-\rho^2}\\
\gamma \e^{\ri\delta} +\ri \rho \mp \sqrt{\gamma^2-\rho^2}
\end{array}
\right),
\end{equation}
where $n_\pm$ is the normalization constant. (The little bit complicated choice of eigenvectors here is to avoid the accidental degeneracy when $\rho = \pm \gamma \sin\delta$.) It can be checked that $|E_\pm\>$ are also eigenvectors of $\cP_0\cT$ with the eigenvalues to have modulus one.
\begin{equation}
\cP_0\cT|E_\pm\> = \e^{\ri \alpha_\pm} |E_\pm\>.
\end{equation}
We can always choose the phases of the eigenvectors such that the eigenvalues of $\cP_0\cT$ to be one.

Solving the self-adjoint condition in Eq.~\eqn{eqn:self-adjoint}, we get the form of the metric operator,
\ba{}
W_0 &=& u\left[\gamma\sigma_0 + (0, v\sin\delta - \rho\cos\delta , v\cos\delta + \rho\sin\delta)\cdot\bm{\sigma}\right] \nn
&=& u \left(
\begin{array}{cc}
\gamma + (v\cos\delta + \rho \sin\delta) & -\ri (v\sin\delta - \rho\cos\delta )\\
\ri (v\sin\delta - \rho\cos\delta) & \gamma - (v\cos\delta + \rho \sin\delta)
\end{array}
\right),
\ea
where $u$ and $v$ are arbitrary real constants with the constraints $u\gamma>0$ and $v^2<\gamma^2-\rho^2$. The eigenvalues of $W_0$ are
\begin{equation}
\omega_\pm = u \left(\gamma \pm \sqrt{\rho^2+v^2} \right).
\end{equation}
They are positive definite. With this metric operator, the eigenvectors of $H_0$ are orthogonal
\begin{equation}
\<E_+|W_0|E_-\>=0=\<E_-|W_0|E_+\>
\end{equation}
and normalized to
\begin{equation}
\cN_\pm \equiv \<E_\pm|W_0|E_\pm\> =
4 |n_\pm|^2 u \gamma \sqrt{\gamma^2-\rho^2} \left(\sqrt{\gamma^2-\rho^2} \pm v \right).
\end{equation}
Given $u$ and $v$, one can always normalize the eigenvector by choosing proper $n_\pm$. It is also true in the reversed way, for arbitrary non-vanishing $n_\pm$, one can always tune $u$ and $v$ in the metric to normalized the eigenvectors.

A generic matrix in the coset of $\dsp\frac{\rGL(2,\mathbb{R})}{\rGL(1,\mathbb{R}) \times \rGL(1,\mathbb{R})}$ can be parameterized either by
\begin{equation}
R_1(\theta,\varphi) = \e^{-\varphi \sigma_3 /2} \e^{-\ri \theta \sigma_2 /2},
\end{equation}
or by
\begin{equation}
R_2(\theta,\varphi) = \e^{-\varphi \sigma_3 /2} \e^{-\theta \sigma_1 /2}.
\end{equation}
Note that they are not equivalent. 
According to these transformations, we have the general \cP\ has the form either as
\begin{equation}
\cP_1 = R_1(\theta,\varphi) \cP_0 R_1^{-1}(\theta,\varphi) =
\left(
\begin{array}{cc}
\cos\theta & \e^{-\varphi}\sin\theta \\
\e^{\varphi} \sin\theta & - \cos\theta
\end{array}
\right),
\label{eqn:P1}
\end{equation}
or
\begin{equation}
\cP_2 = R_2(\theta,\varphi) \cP_0 R_2^{-1}(\theta,\varphi) =
\left(
\begin{array}{cc}
\cosh\theta & \e^{-\varphi}\sinh\theta \\
-\e^{\varphi}\sinh\theta  & - \cosh\theta
\end{array}
\right).
\end{equation}
Two types of \cP\cT-symmetric Hamiltonians can be obtained by the corresponding transformations,
\begin{eqnarray}
H_1 &=& R_1(\theta,\varphi) H_0 R_1^{-1}(\theta,\varphi)\nn
&=& \left(
\begin{array}{cc}
e + \gamma\cos\delta\cos\theta - \ri\rho\sin\theta &
(\gamma\cos\delta\sin\theta - \ri\gamma\sin\delta + \ri\rho\cos\theta) \e^{-\varphi}\\
(\gamma\cos\delta\sin\theta + \ri\gamma\sin\delta + \ri\rho\cos\theta) \e^{\varphi} &
e - \gamma\cos\delta\cos\theta + \ri\rho\sin\theta
\end{array}
\right),
\label{eqn:H1}\\
H_2 &=& R_2(\theta,\varphi) H_0 R_2^{-1}(\theta,\varphi)\nn
&=& \left(
\begin{array}{cc}
e + \gamma\cos(\delta+\ri\theta)
& -\ri [\gamma\sin(\delta+\ri\theta) - \rho] \e^{-\varphi}\\
\ri [\gamma\sin(\delta+\ri\theta) + \rho] \e^{\varphi}
& e - \gamma\cos(\delta+\ri\theta)
\end{array}
\right).
\label{eqn:H2}
\end{eqnarray}
Finally, the metric operators for each $H$ transformed accordingly
\begin{eqnarray}
W_1 &=& \left( R_1^{-1} \right)^\dag (\theta,\varphi) W_0 R_1^{-1}(\theta,\varphi)\nn
&=& u
\left(
\begin{array}{cc}
\left[\gamma + \cos\theta (\rho\sin\delta + v\cos\delta) \right]\e^\varphi &
\sin\theta(\rho\sin\delta + v\cos\delta) + \ri(\rho\cos\delta - v\sin\delta)\\
\sin\theta(\rho\sin\delta + v\cos\delta) - \ri (\rho\cos\delta - v\sin\delta) &
 \left[\gamma - \cos\theta (\rho\sin\delta + v\cos\delta) \right] \e^{-\varphi}
\end{array}
\right),\nn
\\
W_2 &=& \left( R_2^{-1} \right)^\dag (\theta,\varphi) W_0 R_2^{-1}(\theta,\varphi)\nn
&=& u
\left(
\begin{array}{cc}
\left[\gamma\cosh\theta + (\rho\sin\delta + v\cos\delta) \right]\e^\varphi &
\gamma\sinh\theta + \ri(\rho\cos\delta - v\sin\delta)\\
\gamma\sinh\theta - \ri (\rho\cos\delta - v\sin\delta) &
\left[\gamma\cosh\theta - (\rho\sin\delta + v\cos\delta) \right] \e^{-\varphi}
\end{array}
\right).
\end{eqnarray}

Now let us see what goes wrong when a Jordan block is formed. Remember that similarity transformations do not change the diagonalizability of a matrix. Let us work in $\cP_0$ frame. The matrix $H_0$ in Eq.~(\ref{eqn:P0T-H}) forms a Jordan block if and only if $\gamma^2=\rho^2\neq 0$. Because $v^2<\gamma^2-\rho^2$, to avoid double-limit, we fix $v=0$. Let us define a small positive parameter $\epsilon$ by
\begin{equation}
\gamma^2 \equiv \rho^2 (1-\epsilon).
\end{equation}
For arbitrary value of  $u$, when $\epsilon$ is small, the larger eigenvalue of $W_0$, $\omega_>$ behaves well but the smaller one, $\omega_<$ is vanishing,
\begin{equation}
\omega_> \sim 2 u \gamma, \qquad {\rm and} \qquad \omega_< \sim \half u \gamma \epsilon.
\end{equation}
A similar ill behavior can be found in the normalization of eigenvectors of $H_0$,
\begin{equation}
\cN_\pm \sim 4 |n_\pm|^2 u r^3 \epsilon.
\end{equation}
When $H_0$ has a Jordan block structure, one can only find one eigenvector. Another vector to span the linear space lives in the Jordan chain.
The non-diagonalizable $H_0$ has the form
\begin{equation}
H_0 = \left(
\begin{array}{cc}
e + \gamma \cos\delta & \ri \gamma (1 - \sin\delta)\\
\ri\gamma (1 + \sin\delta ) & e - \gamma\cos\delta
\end{array}
\right), \quad {\rm with}\quad \gamma\neq 0.
\end{equation}
The eigenvalue of $H_0$ is $e$ and the only eigenvector is
\begin{equation}
|\Phi_0\> = n_0 \left(
\begin{array}{c}
1-\sin\delta \\
\ri\cos\delta
\end{array}
\right),
\end{equation}
where $n_0$ is the normalization constant. It is obvious that $|\Phi_0\>$ is an eigenvector of \cP\cT,
\begin{equation}
\cP\cT|\Phi_0\> = \frac{n_0^*}{n_0} |\Phi_0\>.
\end{equation}
The second vector in the Jordan chain satisfies
\begin{equation}
(H_0 - e\sigma_0 ) |\Phi_1\> = |\Phi_0\>.
\end{equation}
The solution is
\begin{equation}
 |\Phi_1\> = n_0
\left(
\begin{array}{c}
\frac{1-\sin\delta}{\gamma \cos\delta} \\
0
\end{array}
\right)
+ \alpha  |\Phi_0\>
\end{equation}
with $\alpha$ to be an arbitrary constant. When $\alpha$ is real, $|\Phi_1\>$ is also an eigenvector of \cP\cT\ with the same eigenvalue as $|\Phi_0\>$,
\begin{equation}
\cP\cT|\Phi_1\> = \frac{n_0^*}{n_0} |\Phi_1\>.
\end{equation}

%
%

\section{Pseudo-Hermiticity}
\label{sec:P-pseudo}

In this section, we explore the so called pseudo-Hermitian matrices, an alternative formulation of non-Hermitian matrices with possibly all real eigenvalues. Let us start with a Hermitian involutory operator \tP :
\be{eqn:P-pseudo-P}
\tP=\tP^\dag, \qquad {\rm and} \qquad \tP^2=\one.
\ee
A \tP-pseudo-Hermitian matrix Hamiltonian \tH\ satisfies
\be{}
\tP \tH \tP = \tH^\dag \quad \Leftrightarrow \quad \tP \tH = \tH^\dag \tP.
\ee
Clearly, in this formulation, the \tP\ operator plays the role of the metric operator except that \tP\ may have negative eigenvalues. In general, the linear vector space with an inner-product defined by an indefinite metric operator is called Krein space. In the case of finite dimensions, it is a Pontrjagin space.

\subsection{Unitary transformations}

A unitary transformation preserves the property for the \tP\ operator defined in Eq.~\eqn{eqn:P-pseudo-P}. Just like the method used in the previous section, we may start with a simple operator $\tP_0$. We then use unitary transformations to obtain all possible \tP\ operator. For the case of $m+n$ dimensions,
\begin{equation}
\tP(m,n) = U \tP_0(m,n) U^\dag, \qquad {\rm with} \quad U\in \rU(m+n).
\label{eqn:PPmn}
\end{equation}
There are $(m+n)^2$ real parameters in $U$. Again, we will see later that not all parameters enter $\tP(m,n)$.

If we can find a $\tP_0$-pseudo-Hermitian matrix Hamiltonian $\tH_0$, then all the \tP-pseudo-Hermitian matrix Hamiltonians \tH\ can be obtained by the same unitary transformation,
\begin{equation}
\tH = U \tH_0 U^\dag.
\end{equation}

Without loss of generality, let us choose the $\tP_0(m,n)$ operator to be the same as the parity operator in the previous section,
\begin{equation}
\tP_0(m,n) = {\rm Diag}\{1,\cdots,1,-1,\cdots,-1\} = \left(
\begin{array}{cc}
\one_{m\times m} & 0\\
0 & -\one_{n\times n}
\end{array}
\right).
\end{equation}
A $\tP_0$-pseudo-Hermitian matrix Hamiltonian has the block form
\begin{equation}
\tH_0 = \left(
\begin{array}{cc}
A_{m\times m} & \ri B_{m\times n}\\
\ri (B_{m\times n})^\dag & D_{n\times n}
\end{array}
\right),
\label{eqn:P0H0}
\end{equation}
with $B$ to be arbitrary and $A$ and $D$ to be Hermitian: $A=A^\dag$ and $D=D^\dag.$ Obviously, $\tH_0$ has $m^2+n^2+2mn=(m+n)^2$ real parameters.

Now, we perform the unitary transformation in \eqn{eqn:PPmn} to get the general \tP\ operator. Note that the little group of $\tP_0(m,n)$ in $\rU(m+n)$ will leave $\tP_0(m,n)$ invariant. This little group is the direct sum of $\rU(m)$ and $\rU(n)$. Thus, among $(m+n)^2$ real parameters in the group of $\rU(m+n)$, $m^2$ in the subgroup of $\rU(m)$ and $n^2$ in the subgroup of $\rU(n)$ do not enter \tP. There are $(m+n)^2-m^2-n^2=2mn$ real parameters in \tP. The coset space which parameterizes the nontrivial similarity transformations of the \tP\ operators is a so called complex Grassmann manifold, or complex Grassmannian,
\ba{eqn:Grassman}
\frac{\rU(m+n)}{\rU(m)\times \rU(n)} = {\rm G}_m(\CC^{m+n})
\ea
As coset space it can be parameterized by the Lie algebra elements of the type
\be{g2}
\ru(m+n)\ominus \left(\ru(m)\oplus \ru(n)\right)=\left\{a\in \ru(m+n)\Big|
a=\left(
\begin{array}{ll}
0 & b\\
-b^ \dd & 0
\end{array}
\right),~ b\in \CC^{m\times n}\right\},
\ee
i.e.~by anti-Hermitian matrices $a=-a^\dd\in \CC^{(m+n)\times (m+n)}$ with the $\ru(m)$ and $\ru(n)$
blocks removed from the block-diagonal. For the corresponding coset space elements $U$ of the group it holds the representation
\ba{g3}
U=\e^{ax}=\left(\begin{array}{ll}
\cos(\sqrt{bb^\dd}x) & b\frac{\sin(\sqrt{b^\dd b}x) }{\sqrt{b^\dd b}}\\
-\frac{\sin(\sqrt{b^\dd b}x) }{\sqrt{b^\dd b}}b^\dd & \cos(\sqrt{b^\dd b}x)
\end{array}\right),\qquad x\in \RR.
\ea

The general \tP-pseudo-Hermitian Hamiltonian can be defined as
\begin{equation}
\tH=U \tH_0 U^\dag.
\end{equation}
It has $(m+n)^2+2mn$ real parameters.

In general $\tH_0$ is not Hermitian because the factor of $\ri$ in front of $B$ and $B^\dag$. To make it Hermitian, one has to choose $B=0$. The Hermitian limit of $\tH_0$ has the block-diagonal form
\begin{equation}
h_0 = \left(
\begin{array}{cc}
A_{m\times m} & 0\\
0 & D_{n\times n}
\end{array}
\right),
\end{equation}
where $A=A^\dag, D=D^\dag.$ It has $m^2+n^2$ real parameters. Using the unitary transformation in Eq.~\eqn{eqn:Grassman}, we get the general Hermitian matrix,
\begin{equation}
h=U h_0 U^\dag.
\end{equation}
Now, the question is whether all $(m+n)\times(m+n)$ Hermitian matrices belong to $h$. They do. Here is the reason. Since $A$ and $D$ are Hermitian, they can be written as unitary transformations of diagonal matrices,
\begin{eqnarray}
A = U_m a_m U_m^\dag \quad &{\rm with}& \quad U_m \in \rU(m),\nn
D = U_n d_n U_n^\dag \quad &{\rm with}& \quad U_n \in \rU(n),\nn
\end{eqnarray}
where $a_m$ and $d_n$ are diagonal matrices. The combination of $U$, $U_m$, and $U_n$ re-ensemble all the members in $\rU(m+n)$. Hence, $h$ includes all Hermitian matrices with the same dimension.

One advantage for \tP-pseudo-Hermiticity is that the \tP\ operator defines a Krein space spanned by the eigenvectors of \tH.
\be{Krein}
(\psi,\phi)_{\tP} \equiv \<\psi|\tP|\phi\>.
\ee
All eigenvectors of \tH\ with different eigenvalues are orthogonal with respect to this inner product. However, the norm is not positive definite. Only if the $\tP$-pseudo-Hermitian matrix Hamiltonian \tH\ has an entire real spectrum and it is diagonalizable, a proper Hilbert space can be defined by a positive definite metric operator $\tW$ as in Eq.~\eqn{eqn:inner}. Not only different eigenvectors of \tH\ are orthogonal respect to this $\tW$-inner-product, but also the eigenvectors can be normalized properly. In general, $\tW$ is dynamic, namely \tH-dependent: $\tW\tH=\tH^\dag \tW.$ $\tW$ transforms the same way when one applies the unitary transformation on a \tP-pseudo-Hermitian matrix Hamiltonian \tH:
\begin{equation}
\tW = U \tW_0 U^\dag.
\end{equation}

\subsection{Jordan blocks}

Not all \tP-pseudo-Hermitian matrices with real eigenvalues are diagonalizable. They may form Jordan blocks. In $N\times N$ space, one can always find a unitary transformation $U$ such that
\begin{equation}
\tP=U\tP_0U^\dag=S_i \oplus S_j \oplus \cdots, \qquad {\rm with} \quad i+j+\cdots=N,
\end{equation}
where $S_n$ is the so called standard involutory permutation (SIP),
\be{sip}
S_n=\left(
      \begin{array}{ccccc}
        0 & 0 & \cdots & 0 & 1 \\
        0 & 0 & \cdots & 1 & 0 \\
        \vdots & \vdots & \iddots & \vdots & \vdots \\
        0 & 1 & \cdots & 0 & 0 \\
        1 & 0 & \cdots & 0 & 0 \\
      \end{array}
    \right),\qquad S_n^2=\one_{n\times n},
\ee
with units on the skew-diagonal. They provide the intertwining metrics for Jordan blocks in Eq.~\eqn{eqn:Jn}. For $\lambda\in \RR$,
\ba{g5}
S_n J_n(\lambda)S_n=J_n^\dd(\lambda).
\ea
Obviously, it holds that
\be{g6}
\Tr(S_{2n})=0,\qquad \Tr(S_{2n+1})=1.
\ee
So that (because of $S_n^2=\one_{n\times n}$) we should have the similarity relations $S_{2n}\cong \tP_0(n,n)$, \ $S_{2n+1}\cong \cP_0(n+1,n)$. [This follows from the invariance of the trace under similarity transformations and the fact that $\Tr(\tP_0(n,n))=0$, $\Tr(\tP_0(n+1,n))=1$.] A simple calculation gives
\ba{g7}
S_{2n}&=&q\tP_0(n,n)q^{-1},\nn
q&=&\frac1{\sqrt2}\left(
\begin{array}{cc}
\one_{n\times n} & -S_n\\
S_n & \one_{n\times n}
\end{array}
\right)=e^{b\pi/4},\quad b=\left(
\begin{array}{cc}
0 & -S_n\\
S_n & 0
\end{array}
\right),\nn
q^{-1}&=&\frac1{\sqrt2}\left(
\begin{array}{cc}
\one_{n\times n} & S_n\\
-S_n & \one_{n\times n}
\end{array}
\right),
\ea
as well as
\ba{g8}
S_{2n+1}&=&q\tP_0(n+1,n)q^{-1},\nn
q&=&\frac1{\sqrt2}\left(
\begin{array}{ccc}
\one_{n\times n} & 0&-S_n\\
0&\sqrt2&0\\
S_n & 0& \one_{n\times n}
\end{array}
\right)= \e^{b\pi/4},\quad b=\left(
\begin{array}{ccc}
0 & 0&-S_n\\
0&0&0\\
S_n & 0&0
\end{array}
\right),\nn
 q^{-1}&=&\frac1{\sqrt2}\left(
\begin{array}{ccc}
\one_{n\times n} & 0&S_n\\
0&\sqrt2&0\\
-S_n & 0& \one_{n\times n}
\end{array}
\right).
\ea
Here we have used the fact that
\ba{g9}
b&=&\left(
\begin{array}{cc}
0 & -S_n\\
S_n & 0
\end{array}
\right)=\left(
\begin{array}{cc}
0 & -1\\
1 & 0
\end{array}
\right)\otimes S_n,\nn
&&\left(
\begin{array}{cc}
0 & -1\\
1 & 0
\end{array}
\right)\left(
\begin{array}{cc}
0 & -1\\
1 & 0
\end{array}
\right)=-\one_{2\times 2},
\ea
i.e. that $\left(
\begin{array}{cc}
0 & -1\\
1 & 0
\end{array}
\right)$ can be considered as 2-dimensional realification of the imaginary unit $\ri$ (a representation of $\ri$ in terms of a real $2\times 2$ matrix). In other words,
we can identify
\be{g9a} \ri=\left(
\begin{array}{cc}
0 & -1\\
1 & 0
\end{array}
\right)
\ee
and calculate
\ba{g10}
\e^{\ri S_n x}=\cos(x)\one_{n\times n} + \ri\sin(x)S_n
\ea
what for $x=\pi/4$ and via Eq.~\rf{g9a} just gives Eq.~\rf{g8}.

It can be verified that the following matrix Hamiltonian is \tP-pseudo-Hermitian,
\begin{equation}
\tH=U\tH_0U^\dag=\tH_i \oplus \tH_j \oplus \cdots, \qquad {\rm with} \quad i+j+\cdots=N,
\end{equation}
where $\tH_n$ has a structure looks like a ``rotated Hermitian matrix'',
\be{rotated-H}
\tH_n=\left(\!\!\!
      \begin{array}{ccccc}
        a_{11} + \ri b_{11} & a_{12} + \ri b_{12} & \cdots & \!\!a_{1,n-1} + \ri b_{1,n-1}\!\!\!\! & a_{1n} \\
        a_{21} + \ri b_{21} & a_{22} + \ri b_{22} & \cdots & \!\!\!\!a_{2,n-1} & a_{1,n-1} - \ri b_{1,n-1}\!\! \\
        \vdots & \vdots & \iddots & \vdots & \vdots \\
        \!\!a_{n-1,1} + \ri b_{n-1,1}\!\!\!\! & a_{n-1,2} & \cdots & a_{22} - \ri b_{22} & a_{12} - \ri b_{12} \\
        a_{n1} & \!\!a_{n-1,1} - \ri b_{n-1,1}\!\! & \cdots & a_{21} - \ri b_{21} & a_{11} - \ri b_{11} \\
      \end{array}
    \!\!\!\right)
\ee
with all $a_{ij}$ and $b_{ij}$ to be real parameters.

The advantage for this parametrization is that the Jordan block can easily be constructed. If we set
$a_{11}=a_{22}=\cdots=\lambda$, $a_{12}=a_{23}=\cdots=1$, and all others to be zero, $\tH_n$ becomes a Jordan block of order $n$. Thus, we have shown that all possible Jordan blocks emerge in \tP-pseudo-Hermitian matrices. Of course, the construction here is only a sufficient condition to form Jordan blocks, it is not necessary.

When Jordan block is formed, $\tH$ is no longer diagonalizable and the metric operator cannot be properly defined. We will show explicit examples in $2\times 2$ later.

\subsection{Relation between \cP\cT\ symmetry and pseudo-Hermiticity}

In the earlier papers about \cP\cT-symmetric matrix Hamiltonians, only real-symmetric parity operators are considered \cite{BBJ,BMW,BenderReview}. In this case, the parity operator in \cP\cT\ symmetry is the same as the \cP\ operator in \cP-pseudo-Hermiticity. The transformations between different real-symmetric \cP\ operators are naturally orthogonal, which can be considered as a special case of both unitary and real-similarity transformations. As a consequence, only (complex) symmetric Hamiltonians are constructed. Thus, the \cP\cT\ symmetry condition and \cP-pseudo-Hermiticity condition coincide,
\be{}
\cP H = H^* \cP \quad \Leftrightarrow \quad \cP H = H^\dag \cP.
\ee
Although this limits the number of parameters in the Hamiltonian matrices, it has the advantage to have both the \cP\cT\ eigenstates {\it and}\ \cP\ being an indefinite metric operator. To find the proper metric $W$, one may construct the so-called \cC\ operator which commutes with both $H$ and \cP\cT\ \cite{BBJ}. By construction, the eigenvalues of \cC\ are set to cancel the negative signs in the \cP-norm. In this case, the metric operator can be chosen as $W=\cP\cC.$

If we look at the relation between two formulations from a different angle: For a $\cP\cT$-symmetric matrix Hamiltonian $H$, can we find a $\tP$ operator ($\tP\neq\cP$ in general) such that $H$ is also $\tP$-pseudo-Hermitian? Or for a $\tP$-pseudo-Hermitian matrix Hamiltonian $\tH$, can we find a parity $\cP$ such that $\tH$ is also $\cP\cT$-symmetric? These questions are much harder to answer.

To answer these questions, let us recall a theorem from linear algebra: For any given $n\times n$ matrix $B$, there exists a similarity transformation, $A_B$, which transposes $B$. That is, $A_BBA_B^{-1}=B^T$. A simple proof of this theorem can be found in \ref{appendixA}. Equipped with this theorem, it is straight forward to verify that a $\cP\cT$-symmetric $H$ must satisfy
\be{}
\cQ H = H^\dag \cQ \quad {\rm with} \quad \cQ\equiv A_H^*\cP,
\ee
where $A_H$ is the transpose matrix of $H$: $A_HHA_H^{-1}=H^T$. For any given $H$, there are many $A_H$s which can transpose it. Among these $A_H$s, if some $A_H$s make \cQ\ Hermitian, $\cQ=\cQ^\dag$, then $H$ is $\cQ$-pseudo-Hermitian. If one can further find an $A_H$ such that \cQ\ is not only Hermitian but also a involution, $\cQ^2=\one$, then \cQ\ can be considered as the \tP\ operator in \tP-pseudo-Hermiticity discussed in this section. We will show that in the case of $2\times 2$, such a \tP\ operator always exists. That is, any $2\times 2$ $\cP\cT$-symmetric matrix Hamiltonian is always \tP-pseudo-Hermitian for some \tP.

The reverse problem is similar. For a \tP-pseudo-Hermitian matrix Hamiltonian $\tH$, it satisfies
\be{}
\cQ \tH = \tH^* \cQ \quad {\rm with} \quad \cQ\equiv \left(A_H^*\right)^{-1}\tP,
\ee
If one can find an $A_H$ such that $\cQ=\cQ^*$ and $\cQ^2=\one$, then \cQ\ can be used as the parity operator in the previous section. Namely, $\tH$ is \cP\cT-symmetric. Again, in $2\times 2$, this always can be done.

\subsection{$2\times 2$}

In this subsection, we show explicit examples of $2\times 2$ pseudo-Hermitian matrices. We start with $\tP_0$ to be the third Pauli Matrix,
\begin{equation}
\tP_0 = \sigma_3
=\left(
\begin{array}{cc}
1 & 0\\
0 & -1
\end{array}
\right).
\end{equation}
A $\tP_0$-pseudo-Hermitian Hamiltonian has the form
\begin{equation}
\tH_0 = e\sigma_0 + (\ri \rho \sin\delta , \ri \rho \cos\delta, \gamma ) \cdot\bm{\sigma}
= \left(
\begin{array}{cc}
e + \gamma & \rho \e^{\ri\delta}\\
-\rho \e^{-\ri\delta} & e - \gamma
\end{array}
\right) .
\label{eqn:P0pseudo-H}
\end{equation}
The eigenvalues of $\tH_0$ are
\begin{equation}
E_\pm = e \pm \sqrt{\gamma ^2 - \rho^2}.
\end{equation}
They are real when $\gamma^2 \ge \rho^2$. The eigenvectors are
\begin{equation}
|E_\pm\> = n_\pm \left(
\begin{array}{c}
\gamma \pm \sqrt{\gamma^2-\rho^2}\\
-\rho \e^{-\ri\delta}
\end{array}
\right),
\end{equation}
where $n_\pm$ is the normalization constant. Solving the self-adjoint condition in Eq.~\eqn{eqn:self-adjoint}, we get the form of the metric operator,
\begin{equation}
\tW_0 = u \left[ \gamma \sigma_0 + \left(\rho\cos\delta , -\rho\sin\delta ,v \right) \cdot\bm{\sigma} \right]
=
u \left(
\begin{array}{cc}
\gamma + v  & \rho \e^{\ri\delta}\\
\rho \e^{-\ri\delta} & \gamma - v
\end{array}
\right) ,
\end{equation}
where $u$ and $v$ are arbitrary constants with the constraints $u\gamma>0$ and $v^2<\gamma^2-\rho^2$. The eigenvalues of $\tW_0$ are
\begin{equation}
\omega_\pm = u \left(\gamma \pm \sqrt{\rho^2+v^2} \right).
\end{equation}
They are positive definite. With this metric operator, the eigenvectors of $\tH_0$ with different eigenvalues are orthogonal
\begin{equation}
\<E_+|\tW_0|E_-\>=0=\<E_-|\tW_0|E_+\>
\end{equation}
and they are normalized to
\begin{equation}
\cN_\pm \equiv \<E_\pm|\tW_0|E_\pm\> =
2 |n_\pm|^2 u \sqrt{\gamma^2 - \rho^2} \left(\gamma \pm \sqrt{\gamma^2- \rho^2}\right) \left(\sqrt{\gamma^2- \rho^2}\pm v \right).
\end{equation}
Given $u$ and $v$, one can always normalize the eigenvector by choosing  $n_\pm$ properly. It is also true in the reversed way, for arbitrary non-vanishing $n_\pm$, one can always tune $u$ and $v$ in the metric to normalized the eigenvectors.

A generic unitary matrix in the coset $G_1(\CC\,^2)$ can be parameterized as
\begin{equation}
U(\theta,\varphi) = \e^{-\ri \varphi \sigma_3 /2} \e^{-\ri \theta \sigma_2 /2} .
\end{equation}
It transforms $\tP_0$ to
\begin{equation}
\tP =  U(\theta,\varphi) \tP_0 U^\dag(\theta,\varphi)
= {\bf n}^r\cdot\bm{\sigma}
=\left(
\begin{array}{cc}
\cos\theta & \e^{-\ri\varphi}\sin\theta \\
\e^{\ri\varphi}\sin\theta  & - \cos\theta
\end{array}
\right),
\end{equation}
where ${\bf n}^r\equiv (\sin\theta\cos\varphi,\sin\theta\sin\varphi,\cos\theta)$ is a unit vector. The \tP-pseudo-Hermitian Hamiltonian can be obtained by the same unitary transformation,
\begin{equation}
\tH= U(\theta,\varphi) \tH_0 U^\dag(\theta,\varphi) = e\sigma_0 + \left( \gamma\, {\bf n}^r + \ri\rho\sin\delta\, {\bf n}^\theta + \ri\rho\cos\delta\, {\bf n}^\varphi \right)\cdot\bm{\sigma},
\label{eqn:P-pseudo-H}
\end{equation}
where ${\bf n}^\theta \equiv (\cos\theta\cos\varphi,\cos\theta\sin\varphi,-\sin\theta)$ and ${\bf n}^\varphi \equiv (-\sin\varphi,\cos\varphi,0)$ are two unit vectors perpendicular to ${\bf n}^r$. Finally, the metric operator for this $\tH$ is
\begin{equation}
\tW= U(\theta,\varphi) \tW_0 U^\dag(\theta,\varphi) = u \left[ \gamma \sigma_0 + \left( v\, {\bf n}^r +\rho\cos\delta\, {\bf n}^\theta - \rho\sin\delta\, {\bf n}^\varphi \right)\cdot\bm{\sigma} \right].
\end{equation}

Now let us see what goes wrong when we approach a Jordan block. Unitary transformations do not change the diagonalizability of a matrix. For simplicity, we work in $\tP_0$ frame. $\tH_0$ forms Jordan block if and only if $\gamma^2=\rho^2 \neq 0.$ Define a small positive parameter $\epsilon$ by
\begin{equation}
\rho^2 \equiv \gamma^2 (1-\epsilon).
\end{equation}
To avoid double-limit, we set $v=0$. For a fixed $u$, when $\epsilon$ is small, the larger eigenvalue of $\tW_0$, $\omega_>$ behave well but the smaller one, $\omega_<$ is vanishing,
\begin{equation}
\omega_> \sim 2 u\gamma, \qquad {\rm and} \qquad \omega_< \sim \half u\gamma  \epsilon.
\end{equation}
A similar ill behavior can be found in the normalization of eigenvectors of $\tH_0$,
\begin{equation}
\cN_\pm \sim 2 |n_\pm|^2 u \gamma^3 \epsilon.
\end{equation}

When $\tH_0$ is a Jordan block, one can only find one eigenvector. Another vector lives in the Jordan chain. For simplicity, let us assume $\gamma=\rho\neq 0.$ In this case, $\tH_0$ has the form
\begin{equation}
\tH_0 = \left(
\begin{array}{cc}
e + \gamma & \gamma \e^{\ri\delta}\\
-\gamma \e^{-\ri\delta} & e - \gamma
\end{array}
\right).
\end{equation}
The eigenvalue of $\tH_0$ is $e$ and the only eigenvector is
\begin{equation}
|\Phi_0\> = n_0 \left(
\begin{array}{c}
1 \\
-\e^{-\ri\delta}
\end{array}
\right),
\end{equation}
where $n_0$ is the normalization constant. The second vector in the Jordan chain satisfies
\begin{equation}
(\tH_0 - e\sigma_0 ) |\Phi_1\> = |\Phi_0\>.
\end{equation}
The solution is easy to find,
\begin{equation}
 |\Phi_1\> = n_0
\left(
\begin{array}{c}
0 \\
\frac{1}{\gamma}\e^{-\ri\delta}
\end{array}
\right)
+ \alpha  |\Phi_0\>,
\end{equation}
with $\alpha$ to be an arbitrary constant.

All $2\times 2$ \cP\cT-symmetric matrix Hamiltonians are also pseudo-Hermitian with respect to some \tP\ operators and vice versa. For example, the $\cP_0\cT$-symmetric $H_0$ in Eq.~\eqn{eqn:P0T-H} is \tP-pseudo-Hermitian with
\be{}
\tP=(0,\sin\delta,\cos\delta)\cdot\bm{\sigma} = \left(
\begin{array}{cc}
\cos\delta & -\ri\sin\delta \\
\ri\sin\delta  & - \cos\delta
\end{array}
\right).
\ee
The $\tP_0$-pseudo-Hermitian $\tH_0$ in Eq.~\eqn{eqn:P0pseudo-H} is also \cP\cT-symmetric with
\be{}
\cP= \frac{1}{\sqrt{\gamma^2-\rho^2\cos^2\delta}} \left(
\begin{array}{cc}
\gamma & \rho\cos\delta \\
-\rho\cos\delta  & - \gamma
\end{array}
\right).
\ee
One more example, the parity operator considered in Ref.~\cite{BMW} is the special case of $\cP_1$ in \Eqn{eqn:P1} with $\varphi=0$:
\begin{equation}
\cP_1 = R_1(\theta,0) \cP_0 R_1^{-1}(\theta,0) =
\left(
\begin{array}{cc}
\cos\theta & \sin\theta \\
\sin\theta & - \cos\theta
\end{array}
\right).
\end{equation}
The $\cP_1\cT$-symmetric matrix takes the form of $H_1$ in \Eqn{eqn:H1} with $\varphi=0$. This $H_1$ is \tP-pseudo-Hermitian with
\be{}
\tP= \left(
\begin{array}{cc}
\cos\delta\cos\theta & \cos\delta\sin\theta -\ri \sin\delta \\
\cos\delta\sin\theta + \ri \sin\delta & - \cos\delta\cos\theta
\end{array}
\right).
\ee
In Appendix \ref{appendix}, we give the explicit form of \tP\ for a generic $2\times 2$ \cP\cT-symmetric matrices in \Eqn{eqn:H1} and \Eqn{eqn:H2} to be \tP-pseudo-Hermitian. We also give the explicit form of the parity operator \cP\ for a generic $2\times 2$ \tP-pseudo-Hermitian matrices in \Eqn{eqn:P-pseudo-H} to be  \cP\cT-symmetric.

\section{Generalized \cP\cT\ symmetry}
\label{sec:PT-gen}

We generalize \cP\cT\ symmetry by defining only the combined \cP\cT\ operation as an anti-unitary involutionary operator:
\be{}
\bP\cT \equiv  \bP*, \quad {\rm and} \quad (\bP\cT)^2 = \one,
\ee
where $*$ represents the complex conjugation just as in Sec.~\ref{sec:PT}. These conditions immediately leads to
\begin{equation}
\bP\bP^*=\one.
\label{eqn:DDstar}
\end{equation}
Comparing to \cP\cT\ symmetry, the generalized \cP\cT\ symmetry lifts the restriction of the parity operator to be an involution as well as removes the commuting relation between the parity and the time reversal operators. Meanwhile, we maintain the combined operation to be an involution. As shown in Sec.~\ref{sec:PT}, the combined involutory condition and the anti-linear property are what essentially lead to a possible real spectrum and the phenomena of spontaneous \cP\cT\ symmetry breaking. A generalized \bP\cT-symmetric matrix Hamiltonian satisfies
\be{eqn:gen-PT}
[\bH,\bP\cT]=0 \qquad \Leftrightarrow \qquad \bP \bH^* = \bH \bP.
\ee

The transformations on the $\bP$ matrix that preserves \Eqn{eqn:DDstar} are
\be{}
\bP = \Lambda \bP_0 \left( \Lambda^{-1} \right)^*.
\ee
Under this transformation, a generalized $\bP_0\cT$-symmetric matrix Hamiltonian $\bH_0$ transforms to a generalized $\bP\cT$-symmetric matrix Hamiltonian $\bH$ if
\be{eqn:LambdaH}
\bH = \Lambda \bH_0 \Lambda^{-1},
\ee
which turns out to be a similarity transformation. There is no limitation on the transformation matrix $\Lambda$ except it must be invertible. For the case of $N\times N$, $\Lambda \in {\rm GL}(N,\CC).$

When the eigenvalues of $\bH$ are real, the argument in \Eqn{eqn:PTeigen} still holds. That is, the eigenstates of $\bH$ are also eigenstates of $\bP\cT$ with the eigenvalues can be chosen to be one.

If $\bH_0$ not only has an entire real spectrum, but also is diagonalizable, one may solve the self-adjoint condition in \Eqn{eqn:self-adjoint} to find the proper metric operator $\bW_0$. The corresponding metric for $\bH$ can be obtained by
\be{eqn:LambdaW}
\bW = \left( \Lambda^{-1}\right)^\dag \bW_0 \Lambda^{-1}.
\ee

The similarity transformation in \Eqn{eqn:LambdaH} also means that all self-adjoint matrices are generalized \bP\cT-symmetric. There are two ways to see it. The simpler way is to start with a diagonal form of $\bH_0$. It is obviously generalized $\bP_0\cT$-symmetric with $\bP_0=\one.$ The similarity transformation brings $\bH$ to be an arbitrary self-adjoint matrices with the same spectrum. This $\bH$ is generalized $\bP\cT$-symmetric with $\bP = \Lambda \left( \Lambda^{-1} \right)^*.$ Note that although $\bP_0$ is the unity, \bP\ is non-trivial in general: $\bP\neq\one.$

Another way is to start with a diagonal metric operator $W_D$:
\begin{equation}
W_D = {\rm Diag}\{\omega_1,\omega_2,\omega_3,\cdots\}.
\end{equation}
The self-adjoint condition in \Eqn{eqn:self-adjoint} leads to the Hamiltonian with the form
\begin{equation}
H_D=\left(
\begin{array}{cccc}
a_{11} & \frac{2\omega_2}{\omega_1+\omega_2} (a_{12}+\ri b_{12}) & \frac{2\omega_3}{\omega_1+\omega_3} (a_{13}+\ri b_{13}) & \cdots \\
\frac{2\omega_1}{\omega_1+\omega_2} (a_{12}-\ri b_{12}) & a_{22} & \frac{2\omega_3}{\omega_2+\omega_3} (a_{23}+\ri b_{23}) & \cdots\\
\frac{2\omega_1}{\omega_1+\omega_3} (a_{13}-\ri b_{13}) & \frac{2\omega_2}{\omega_2+\omega_3} (a_{23}-\ri b_{23}) & a_{33} &\cdots\\
\vdots & \vdots & \vdots & \ddots
\end{array}
\right)
\label{eqn:W0H0}
\end{equation}
with all $a_{ij}$ and $b_{ij}$ to be real. We observe three properties for a matrix to be self-adjoint respect to a diagonal metric: (i) All the diagonal elements must be real,
\be{}
H_D^{ii} = \left( H_D^{ii} \right)^*;
\ee
(ii) The off-diagonal elements have opposite phases comparing to the mirror element on the other side of the diagonal,
\be{}
H_D^{ij}H_D^{ji} = \left( H_D^{ij}H_D^{ji} \right)^*;
\ee
(iii) And the ratio between two moduli of mirroring off-diagonal elements is determined by the eigenvalues of $W_D$,
\be{eqn:H0_3}
\left| \frac{H_D^{ij}}{H_D^{ji}} \right| = \frac{\omega_j}{\omega_i}.
\ee
Note that $H_D$ is a Hermitian matrix if and only if $W_D$ proportional to the unity matrix.

Now let us see what a generalized $\bP_0\cT$-symmetric matrix with a diagonal $\bP_0$ looks like. Because of \Eqn{eqn:DDstar}, a diagonal $\bP_0$ has matrix elements to be a pure phase on the diagonal:
\begin{equation}
\bP_0 = {\rm Diag}\{\e^{\ri\alpha_1},\e^{\ri\alpha_2},\e^{\ri\alpha_3},\cdots\}.
\end{equation}
Solving the generalized \bP\cT\ symmetry condition in \Eqn{eqn:gen-PT}, we get the matrix elements of $\bH_0$ satisfies
\be{}
h_{ij} \e^{\ri\alpha_j} = h^*_{ij} \e^{\ri\alpha_i}.
\ee
Multiply $\e^{-\ri(\alpha_i+\alpha_j)/2}$ on the both sides, we have
\be{}
h_{ij} \e^{-\ri(\alpha_i - \alpha_j)/2} = \left( h_{ij} \e^{-\ri(\alpha_i - \alpha_j)/2} \right)^*.
\ee
This means that the phase of the $ij$ element of $\bH_0$ is $(\alpha_i - \alpha_j)/2$. Thus $\bH_0$ has the form
\begin{equation}
\bH_0=\left(
\begin{array}{cccc}
r_{11} & r_{12} \e^{\ri(\alpha_1 - \alpha_2)/2} & r_{13} \e^{\ri(\alpha_1 - \alpha_3)/2} & \cdots \\
r_{21} \e^{\ri(\alpha_2 - \alpha_1)/2} & r_{22} & r_{23} \e^{\ri(\alpha_2 - \alpha_3)/2} & \cdots\\
r_{31} \e^{\ri(\alpha_3 - \alpha_1)/2} & r_{32} \e^{\ri(\alpha_3 - \alpha_2)/2} & r_{33} &\cdots\\
\vdots & \vdots & \vdots & \ddots
\end{array}
\right),
\label{eqn:D0H0}
\end{equation}
with $r_{ij}$ real. The form of the generalized $\bP_0\cT$-symmetric $\bH_0$ is very similar to the self-adjoint $H_D$ in \Eqn{eqn:W0H0}. But $\bH_0$ in \Eqn{eqn:D0H0} is more general. The subtle difference lies in the third property of $H_D$ in \Eqn{eqn:H0_3}: $H_D^{ij}$ and $H_D^{ji}$ must be both zero or both non-vanishing. On the contrary, one can take $\bH_0^{ij} = 0$ at the same time $\bH_0^{ji} \neq 0$ for $i\neq j$. In addition, the eigenvalues of $H_D$ must be real and $\bH_0$ allows both real and complex conjugate pair eigenvalues. We conclude that $\bH_0$ is more general than $H_D$, and $H_D$ is a special case of $\bH_0$. Finally, the transformations in \Eqn{eqn:LambdaH} gives us that all self-adjoint matrices process a generalized \bP\cT\ symmetry.

It is very easy to count how many real parameters in an arbitrary $N\times N$ self-adjoint matrix. A generic $N\times N$ complex matrix has $2N^2$ real parameters. The reality of the eigenvalue poses $N$ constraints. So an $N\times N$ self-adjoint matrix has $2N^2-N$ real parameters.

Although all self-adjoint matrices form a subset of generalized \bP\cT-symmetric matrices, the latter have the same number of parameters. It is simply due to the anti-linear involutory property of \bP\cT, a generalized \bP\cT-symmetric matrix have real or complex pair eigenvalues. For an $N\times N$ matrix, it poses $N$ constraints.

Reducing the generalized \bP\cT\ symmetry to the \cP\cT\ symmetry discussed in Sec.~\ref{sec:PT} is trivial, all we need is to start with a real $\bP_0=\cP_0$ and transform only by real matrices. In this way, the resulting parity operator \cP\ is always real and $\bH$ is \cP\cT-symmetric. On the contrary, reducing the
generalized \bP\cT\ symmetry to pseudo-Hermiticity is much more complicated. For a generalized \bP\cT-symmetric matrix $\bH$, $\bP \bH^* = \bH \bP$, if there exist a matrix $A_H$ such that
\be{}
A_H \bH A_H^{-1} =\bH^T \quad {\rm and} \quad A_HA_H^*=\one,
\ee
then $\bH$ satisfies
\be{}
\cQ \bH^\dag = \bH\cQ \quad {\rm with} \quad \cQ\equiv \bP A_H.
\ee
Among these $A_H$s, if some make \cQ\ Hermitian, $\cQ=\cQ^\dag$, then $\bH$ is \cQ-pseudo-Hermitian. Finally, if one can find an $A_H$ such that $\cQ^2=\one$ also holds, then this \cQ\ matrix can be the \tP\ operator discussed in Sec.~\ref{sec:P-pseudo}, that is,  $\bH$ is \tP-pseudo-Hermitian.

For completeness, we give an explicit parametrization of the $2\times 2$ \bP\ operator discussed in this section:
\be{}
\bP= \cosh\varphi\e^{\ri \alpha}\left(
\begin{array}{cc}
\cos\theta + \ri\sin\theta\sin\delta & \ri\left(\sin\theta\cos\delta - \tanh\varphi \right) \\
\ri\left(\sin\theta\cos\delta + \tanh\varphi \right) & \cos\theta - \ri\sin\theta\sin\delta
\end{array}
\right).
\ee
It is straight forward to verify that $\bP\bP^*=\sigma_0$.

\section{Conclusions}
\label{conclusion}

\begin{table}[htb]
\caption{The numbers of real parameters in real-symmetric, Hermitian, \cP\cT-symmetric, \tP-pseudo-Hermitian, self-adjoint, and generalized \bP\cT-symmetric matrices.}
\begin{center}
\begin{tabular}{lcccc}
Dimension & $2$ & $3$ & $m+n$ & $N\gg 1$\\
\hline
\hline
Real-symmetric & $3$ & $6$ & $\frac{1}{2}(m+n)(m+n+1)$ & $\frac{1}{2}N^2$\\
\hline
Hermitian & $4$ & $9$ & $(m+n)^2$ & $N^2$\\
\hline
\begin{tabular}{l}
\cP\cT-symmetric or \\
\tP-pseudo-Hermitian
\end{tabular} & $6$ & $13$ & $(m+n)^2+2mn$ & $N^2 \sim \frac{3}{2}N^2 $ \\
\hline
\begin{tabular}{l}
Self-adjoint or \\
Generalized \bP\cT-symmetric
\end{tabular} & $6$ & $15$ & $2(m+n)^2-(m+n)$ & $2N^2 $ \\
\hline
\hline
\end{tabular}
\end{center}
\label{tab1}
\end{table}

In conclusion, we list the parameter counting of all relevant matrices in Table \ref{tab1}. From the table, in the case of $2\times 2$, \cP\cT-symmetric matrices, \tP-pseudo-Hermitian matrices, generalized \bP\cT-symmetric matrices, and self-adjoint matrices all have the same number of parameters.  In fact, any $2\times 2$ self-adjoint matrix is \cP\cT-symmetric, \tP-pseudo-Hermitian, {\it and} generalized \bP\cT-symmetric. This is not true in another way around because \cP\cT\ symmetry may be broken. Even it does not, a \cP\cT-symmetric matrix or a \tP-pseudo-Hermitian or a generalized \bP\cT-symmetric matrix could form a Jordan block and no longer diagonalizable. In the dimension higher than $2$, a self-adjoint matrices is always generalized \bP\cT-symmetric but not \cP\cT-symmetric or \tP-pseudo-Hermitian in general.

\appendix
\section{Proof of the existence of the transpose matrix}
\label{appendixA}

Suppose that an $n\times n$ matrix $B$ is similar to a Jordan block $J_n$,
\be{}
F B F^{-1} = J_n.
\ee
Consider another $n\times n$ matrix $A_B$ defined as
\be{transpose}
A_B \equiv F^T S_n F,
\ee
where $S_n$ is an SIP defined in Eq.~(\ref{sip}), then $A_B$ is the transpose matrix of $B$,
\be{}
A_BBA_B^{-1}=B^T.
\ee
If matrix $B$ is similar to a direct sum of Jordan blocks,  
\be{}
F B F^{-1} = J_{m_1}\oplus J_{m_2} \oplus\cdots,
\ee
then the transpose matrix can be constructed by replacing $S_n$ in Eq.~(\ref{transpose}) by a similar sum, $S_{m_1}\oplus S_{m_2} \oplus\cdots$. Since any square matrix is either similar to a Jordan block or a direct sum of Jordan blocks, there always exists a transpose matrix as constructed above.

\section{}
\label{appendix}
In this appendix, we give the explicit forms in $2\times 2$ about the equivalence of \cP\cT\ symmetry and \tP\ pseudo-Hermiticity.

The generic \tP-pseudo-Hermitian $\tH$ in \Eqn{eqn:P-pseudo-H} is \cP\cT-symmetric respect to
\begin{equation}
\cP= \pm \frac{1}{\sqrt\Delta_1}\left(
\begin{array}{cc}
\gamma\cos\theta & \cP_{12} \\
\cP_{21} & -\gamma\cos\theta
\end{array}
\right)
\nonumber
\end{equation}
with
\begin{eqnarray*}
\cP_{12} &\equiv& \gamma\sin\theta\cos\varphi + \rho\cos\delta\cos\varphi + \rho\sin\delta\cos\theta\sin\varphi\\
\cP_{21} &\equiv& \gamma\sin\theta\cos\varphi - \rho\cos\delta\cos\varphi - \rho\sin\delta\cos\theta\sin\varphi \\
\Delta_1 &\equiv& \gamma^2\left(\cos\delta\cos\theta\sin\varphi - \sin\delta\cos\varphi\right)^2  + \left(\gamma^2-\rho^2\right)\left(\sin\delta\cos\theta\sin\varphi + \cos\delta\cos\varphi \right)^2.
\end{eqnarray*}

The generic $\cP_1\cT$-symmetric $H_1$ in \Eqn{eqn:H1} is $\tP$-pseudo-Hermitian respect to the Hermitian involution operator
\begin{equation}
\tP= \pm \frac{1}{\sqrt\Delta_2}\left(
\begin{array}{cc}
\gamma\cos\delta\cos\theta & \tP_{12}\\
\tP_{21}& - \gamma\cos\delta\cos\theta
\end{array}
\right)
\nonumber
\end{equation}
with
\begin{eqnarray*}
\tP_{12} &\equiv& \gamma\cos\delta\sin\theta\cosh\varphi - \ri\gamma\sin\delta\cosh\varphi - \ri\rho\cos\theta\sinh\varphi\\
\tP_{21} &\equiv& \gamma\cos\delta\sin\theta\cosh\varphi + \ri\gamma\sin\delta\cosh\varphi + \ri\rho\cos\theta\sinh\varphi\\
\Delta_2 &\equiv& \gamma^2 \cos^2\delta \left(\sin^2\theta\cosh^2\varphi + \cos^2\theta\right) + \left(\gamma\sin\delta\cosh\varphi + \rho\cos\theta\sinh\varphi\right)^2.
\end{eqnarray*}

The generic $\cP_2\cT$-symmetric $H_2$ in \Eqn{eqn:H2} is $\tP$-pseudo-Hermitian with respect to the Hermitian involution operator
\begin{equation}
\tP'= \pm \frac{1}{\sqrt\Delta_3}\left(
\begin{array}{cc}
\gamma\cos\delta\cosh\theta & \tP'_{12}\\
\tP'_{21} & - \gamma\cos\delta\cosh\theta
\end{array}
\right)
\nonumber
\end{equation}
with
\begin{eqnarray*}
\tP'_{12} &\equiv& -\gamma\cos\delta\sinh\theta\sinh\varphi - \ri\gamma\sin\delta\cosh\theta\cosh\varphi - \ri\rho\sinh\varphi\\
\tP'_{21} &\equiv& -\gamma\cos\delta\sinh\theta\sinh\varphi + \ri\gamma\sin\delta\cosh\theta\cosh\varphi + \ri\rho\sinh\varphi \\
\Delta_3 &\equiv& \gamma^2 \cos^2\delta \left(1+\sinh^2\theta\cosh^2\varphi\right) + \left(\gamma\sin\delta\cosh\theta\cosh\varphi + \rho\sinh\varphi\right)^2.
\end{eqnarray*}

\section*{Acknowledgements}
QhW would like to thank Prof.~Jiangbin Gong for helpful discussions.


\begin{thebibliography}{99}

\bibitem{BB}
C.M.~Bender and S.~Boettcher,
``{\it Real Spectra in Non-Hermitian Hamiltonians Having \cP\cT\ Symmetry},''
Phys.~Rev.~Lett.~{\bf 80}, 5243-5246 (1998).

\bibitem{BenderReview}
C.M.~Bender,
Rep.~Prog.~Phys.~{\bf 70}, 947 (2007).

\bibitem{AliReview}
A.~Mostafazadeh,
``{\it Pseudo-Hermitian Representation of Quantum Mechanics},''
Int.~J.~Geom.~Meth.~Mod.~Phys.~{\bf 7}, 1191-1306 (2010).

\bibitem{theory1}Z.H.~Musslimani, K.G.~Makris, R.~El-Ganainy, and  D.N.~Christodoulides, Phys.~Rev.~Lett.~{\bf 100}, 030402 (2008)

\bibitem{theory2}
K.G.~Makris, R.~El-Ganainy, D.N.~Christodoulides, and Z.H.~Musslimani,
``Beam Dynamics in \cP\cT~Symmetric Optical Lattices,''
{\it Phys.~Rev.~Lett.}~{\bf 100}, 103904 (2008).

\bibitem{theory3}
S.~Longhi,
``Bloch Oscillations in Complex Crystals with \cP\cT~Symmetry,''
{\it Phys.~Rev.~Lett.}~{\bf 103}, 123601 (2009);
``Optical Realization of Relativistic Non-Hermitian Quantum Mechanics.''
{\it ibid.}~{\bf 105}, 013903 (2010).

\bibitem{theory4}
C.T.~West, T.~Kottos, and T.~Prosen, {\it Phys.~Rev.~Lett.}~{\bf 104}, 054102 (2010).

\bibitem{exp1}
A.~Guo, G.J.~Salamo, D.~Duchesne, R.~Morandotti, M.~Volatier-Ravat, V.~Aimez, G.A.~Siviloglou, and D.N.~Christodoulides,
``Observation of \cP\cT-Symmetry Breaking in Complex Optical Potentials,''
{\it Phys.~Rev.~Lett.}~{\bf 103}, 093902 (2009).

\bibitem{NaturePhys1}
T.~Kottos,
``Optical physics: Broken symmetry makes light work,''
{\it Nature Physics}~{\bf 6}, 166 (2010).

\bibitem{NaturePhys2}
C.E.~R\"uter, K.G.~Makris, R.~El-Ganainy, D.N.~Christodoulides, M.~Segev, and D.~Kip,
``Observation of parity-time symmetry in optics,''
{\it Nature Physics}~{\bf 6}, 192 (2010).

\bibitem{Science}
L.~Feng, M.~Ayache, J.~Huang, Y.-L.~Xu, M.-H.~Lu, Y.-F.~Chen, Y.~Fainman, A.~Scherer
``Nonreciprocal Light Propagation in a Silicon Photonic Circuit,''
{\it Science}~{\bf 333}, 729 (2011).

\bibitem{Nature}
C.~Bersch , D.N.~Christodoulides , M.-A.~Miri , G.~Onishchukov , U.~Peschel, and A.~Regensburger,
``Parity-time synthetic photonic lattices.''
{\it Nature} {\bf 488}, 7410, 167-171 (2012).

\bibitem{BMW}
C.M.~Bender, P.N.~Meisinger, and Q.~Wang,
``{\it Finite-Dimensional \cP\cT-Symmetric Hamiltonians},''
J.~Phys.~A: Math.~Gen.~{\bf 36}, 6791-6797 (2003) [arXiv:quant-ph/0303174].

\bibitem{WCZ}
Q.-h.~Wang, S.-z,~Chia, and J.-h.~Zhang,
``{\it \cP\cT\ Symmetry as a Generalization of Hermiticity}''
J.~Phys.~A: Math.~Theor.~{\bf 43}, 295301 (2010) [arXiv:1002.2676]

\bibitem{Jordan}
T.F.~Jordan,
{\it Linear Operators for Quantum Mechanics} (John Wiley \& sons, New York, 1969).

\bibitem{BalBook}
L.E.~Ballentine,
{\it Quantum Mechanics: A Modern Development} (World Scientific, Singapore, 1998).

\bibitem{BBJ}
C.M.~Bender, D.C.~Brody, and H.F.~Jones,
``{\it Complex Extension of Quantum Mechanics},''
Phys.~Rev.~Lett.~{\bf 89}, 270401 (2002) [Erratum: {\it ibid.}~{\bf 92} 119902 (2004)].

%



%
%
%


%

\end{thebibliography}
\end{document}